\documentclass[showpacs,preprintnumbers,amsmath,amssymb]{revtex4}

\usepackage[latin1]{inputenc}
\usepackage{makeidx}
\usepackage{layout}        
\usepackage{color}
\usepackage{graphicx}
\usepackage{dcolumn}
\usepackage{bm}

\newcommand{\bra}[1]{\langle #1|}
\newcommand{\ket}[1]{|#1\rangle}

\newcommand{\average}[1]{\langle #1 \rangle}

\begin{document}
\preprint{preprint}

\title{Orbital contribution to the magnetic properties of iron as a function of dimensionality}

\author{ Marie-Catherine Desjonqu\`eres$^*$,  Cyrille Barreteau$^*$, Gabriel Aut\`es$^*$ and Daniel Spanjaard$^{\dag}$ }
\affiliation{$^*$CEA Saclay, DSM/DRECAM/SPCSI, B\^atiment 462, F-91191 Gif sur Yvette, France }

\affiliation{$^{\dag}$Laboratoire de Physique des Solides,
             Universit\'e Paris Sud, B\^atiment 510, F-91405 Orsay, France}

\date{\today}
\begin{abstract}

The orbital contribution to the magnetic properties of Fe in systems of decreasing
dimensionality (bulk, surfaces, wire and free clusters) is investigated using
a tight-binding hamiltonian in an $s, p,$ and $d$ atomic orbital basis set
including spin-orbit coupling and intra-atomic electronic interactions in the
full Hartree-Fock  (HF) scheme, i.e., involving all the matrix elements of the
Coulomb interaction with their exact orbital dependence. Spin and orbital magnetic 
moments and the magnetocrystalline anisotropy energy (MAE) are calculated for
several orientations of the magnetization. The results are systematically compared
with those of simplified hamiltonians which give results close to those 
obtained from the local spin density approximation. The full HF decoupling
leads to much larger orbital moments and MAE which can reach values as large as
1$\mu_B$ and several tens of meV, respectively, in the monatomic wire at the equilibrium
distance. The reliability of the results obtained by adding the so-called 
Orbital Polarization Ansatz (OPA) to the simplified hamiltonians is also
discussed. It is found that when the spin magnetization is saturated the
OPA results for the orbital moment are in qualitative agreement with those
of the full HF model. However there are large discrepancies for the MAE,
especially in clusters. Thus the full HF scheme must be used to investigate
the orbital magnetism and MAE of low dimensional systems.

\end{abstract}

\pacs{75.30.Gw,75.75.+a,75.90.+w}

\maketitle

\section{Introduction}

The magnetic properties of reduced dimensionality systems is an area of growing
interest both for experimentalists and theoreticians. Indeed, the spin magnetic
moment of an atom is strongly dependent on its environment, in particular, on
its coordination number. It usually increases when the latter decreases and may 
even appear in small clusters for some transition metals 
which are not magnetic in the bulk phase \cite{Cox94,Guirado98a,Guirado98b,Barreteau00a}.
Another quantity playing a key role in technological applications, like magnetic recording,
is the magnetocrystalline anisotropy energy (MAE) which is responsible for the tendency of the magnetization
to align along particular directions. It is well known that this MAE is very small in the bulk phase of
ferromagnetic transition metals but may increase by several orders of magnitude
when the dimensionality or the symmetry of the system is reduced. The most striking
specific property of low dimensional systems is perhaps the appearance of a sizable
orbital contribution to the magnetic moment which, on the contrary, is practically
quenched in bulk systems. Obviously, the influence of intra-atomic Coulomb interactions
and spin-orbit coupling, responsible for Hund's rules in the free atom, become
more and more important when the band width due to electron delocalization decreases
and both the spin and orbital moments should tend to their atomic values. Furthermore,
the interest for the orbital moment has been stimulated by a newly acquired physical
technique, the X-ray Magnetic Circular Dichroism (XMCD), which is able to resolve
spin and orbital moments. Such experiments have been carried out by Gambardella {\sl et al.}
\cite{Gambardella02,Gambardella03} who have indeed measured orbital moments as large
as 0.68$\mu_B$ for Co chains on Pt(997) and 1.1$\mu_B$ for a Co adatom on Pt(111),
the corresponding MAE being 2meV and 9meV per Co atom, respectively. Moreover XMCD
experiments carried out on iron clusters\cite{Edmonds00,Binns01} have shown that the
orbital moment is much more enhanced than the spin moment as the size is reduced.

From the theoretical point of view, the spin moment can be obtained in the
Local Spin Density Approximation (LSDA) or with a Stoner-like Tight-Binding
(TB) hamiltonian \cite{Autes06} but the spin orientation is arbitrary. This
is not true when spin-orbit coupling is taken into account and, consequently,
the MAE and the orbital moment no longer vanish. However, in these schemes the 
hamiltonian depends only on the total spin density (LSDA) or spin population
at each site (TB) and not on their repartition between the orbital states.
In other words electronic interactions are averaged which yields underestimated
values of MAE and orbital moments, eventhough these quantities increase when
the dimensionality is reduced\cite{Xie04}. Eriksson {\sl et al.} \cite{Eriksson90} have
proposed to correct this drawback by adding to the total energy a term proportional
to $-\frac{1}{2}\average{{\bf L}}^2$ which will be referred to as the Orbital Polarization
Ansatz (OPA) in the following. This obviously tends to increase $\average{{\bf L}}$ but
is not really justified.

A more rigorous way of obtaining the correct distribution of electrons between
the $d$ orbital states of opposite magnetic quantum numbers $m$ is to take into
account all intra-atomic terms in the Hartree-Fock (HF) decoupling of the
two-body operators representing electron-electron interactions in the hamiltonian
with the exact expression of the matrix elements $U_{\gamma_1\gamma_2\gamma_3\gamma_4}$
of $e^2/|{\bf r}-{\bf r'}|$ as a function of the three Racah parameters A, B and C
relative to $d$ atomic orbitals $\gamma_{\alpha}$. Note that, when using L(S)DA or
a TB hamiltonian parametrized by fits on L(S)DA calculations, some electronic
interactions are implicitly already included. This is usually accounted for
by assuming that on each atom all spin-orbitals (all spin-orbitals with the 
same spin) are equally populated in LDA (LSDA). 

Some attempts
have been made in this direction. However, most often the terms involving three
and four different orbitals have been neglected  which destroys
the rotational invariance of the interaction hamiltonian \cite{Nicolas06} unless appropriate
averages of the matrix elements with two different orbitals are done \cite{Anisimov93,Wierzbowska05}.
Nevertheless there exists in the literature some scarce calculations in which all matrix
elements of the Coulomb interactions were taken into account\cite{Liechtenstein95,Shick04} but, at least to our
knowledge, no systematic study comparing this complete HF scheme to the OPA
has been carried out save for our preliminary work on the Fe monatomic wire\cite{Desjonqueres07} in
the TB approximation with an atomic orbital basis set restricted to $d$ orbitals.
Such a comparison is indeed very interesting since Solovyev {\sl et al.} \cite{Solovyev98}
have shown, in an elegant work, that the OPA cannot be derived analytically from
the full HF hamiltonian except in some very special cases. In the present work
we generalize our previous study\cite{Desjonqueres07}, on the one hand, by using a realistic description
of the electronic states including $sp-d$ hybridization in order to get quantitative results and, on the other hand,
by investigating systems with various dimensionalities (bulk, surfaces, wire, clusters).

The paper is organized as follows. The formalism and the choice of parameters 
for Fe are described in Sec.II. Secs.III and IV
are devoted to bulk and surfaces of bcc iron, respectively. Our results concerning
the monatomic wire and some clusters are given in Secs.V and VI. Finally conclusions
are drawn in Sec.VII.

\section{Formalism}

The hamiltonian of the system is written as:

\begin{equation}
H=H_{TB}+H_{so}+\Delta H_{int}
\label{Htot}
\end{equation}

$H_{TB}$ is a tight-binding hamiltonian parametrized for the non magnetic state
by fitting ab-initio calculations in the LDA (Local Density) or GGA (Generalized
Gradient) approximation, $H_{so}$ is the spin-orbit coupling term and $\Delta H_{int}$ 
describes the change in electronic interactions with respect to $H_{TB}$. 

We choose a non orthogonal basis set made of the real {\it s, p} and {\it d} valence 
atomic orbitals (i.e. cubic harmonics for reasons which will become clear in the
following) centered at each site {\it i}. They are denoted by $\lambda$ and $\mu$
indices ($\lambda, \mu=1,9$) and numbered as follows: $s, p_x, p_y, p_z, d_{xy},
d_{yz}, d_{zx}, d_{x^2-y^2}, d_{3z^2-r^2}$ (the x, y, z coordinates being taken
along the crystal axes) with overlap integrals $S^{\lambda\mu}_{ij}$ depending
on the bonding direction ${\bf R}_{ij}$. The hamiltonian $H_{TB}$ is completely
determined by its intra-atomic matrix elements (i.e., the {\it s, p} and {\it d}
atomic levels) $\epsilon_{i\lambda}$ and its inter-atomic matrix elements (i.e., the
hopping integrals) $\beta^{\lambda\mu}_{ij}({\bf R}_{ij})$. The functions 
$S^{\lambda\mu}_{ij}({\bf R}_{ij})$ and $\beta^{\lambda\mu}_{ij}({\bf R}_{ij})$
are given by the same analytical expressions as a function of two sets (one for
overlap and one for hopping) of ten Slater-Koster (SK) integrals ($ss\sigma, sp\sigma,
sd\sigma, pp\sigma, pp\pi, pd\sigma, pd\pi, dd\sigma, dd\pi, dd\delta$) 
and of the direction cosines of ${\bf R}_{ij}$ \cite{Slater54}. Following
the (MP) scheme developed by Mehl and Papaconstantopoulos \cite{Mehl96}, the 
atomic levels depend on the atomic environment (number of neighbors and
interatomic distances) while the SK integrals are functions of $R_{ij}$ only.
The atomic levels and the SK integrals are written as analytic functions depending
on a number of parameters which are determined by a least mean square fit of the
results of ab-initio LDA or GGA electronic structure (band structure and total
energy) calculations. These parametrizations will be denoted respectively as
TBLDA and TBGGA in the following: the analytical forms of the functions can be
found in Ref.\onlinecite{Mehl96} and the numerical values of the parameters for Fe
in Ref.\onlinecite{Papaweb}.

The spin-orbit coupling hamiltonian $H_{so}$ is given by:

\begin{equation}
H_{so}=\xi {\bf L}.{\bf S}
\label{Hso}
\end{equation}

\noindent
where $ {\bf L}$ and $ {\bf S}$ are the orbital and spin momentum operators,
respectively. Due to the local character of this interaction, only the intra-atomic
matrix elements between {\it d} spin-orbitals have been taken into account. For
more details the reader is referred to Ref.\onlinecite{Autes06} where $\xi$ has
been determined ($\xi=0.06$eV).

Only the intra-atomic electronic interactions are taken into account. For {\it d}
electrons they are written in the Hartree-Fock approximation, i.e.:

\begin{eqnarray}
H_{\text{int},dd}^{\text{HF1}}=& \displaystyle
             \sum_{ \substack{i \gamma_1 \gamma_2 \gamma_3 \gamma_4 \\ \sigma \sigma'} }
             \Big( U_{\gamma_4\gamma_2\gamma_3\gamma_1} \average{n_{i\gamma_3\gamma_4}^{\sigma\sigma}}c_{i \gamma_2\sigma'}^{\dag}  c_{i \gamma_1\sigma'} \nonumber    \\
            &  -     U_{\gamma_4\gamma_2\gamma_1\gamma_3}  \average{n_{i\gamma_3\gamma_4}^{\sigma'\sigma}} c_{i \gamma_2\sigma'}^{\dag}   c_{i \gamma_1\sigma} \Big)
\end{eqnarray}
 
 \noindent 
 as a function of the net density matrix $\average{n_{i\gamma_1\gamma_2}^{\sigma\sigma'}}=\average{c_{i \gamma_2\sigma'}^{\dag} c_{i
 \gamma_1\sigma}}$ when
 all intra-atomic matrix elements of the Coulomb interactions, i.e.:
 
\begin{equation}
U_{\gamma_1\gamma_2\gamma_3\gamma_4}=\bra{\gamma_1(\bm{r}),\gamma_2(\bm{r'})}
\frac{e^2}{|\bm{r-r'}|}\ket{\gamma_3(\bm{r}),\gamma_4(\bm{r'})}
\end{equation}

\noindent
where $\gamma_i$ are a set of atomic {\it d} orbitals, are retained (HF1 model of Ref.\onlinecite{Desjonqueres07}).
These matrix elements can be expressed as a function of the three Racah parameters A, B, C \cite{Griffith61}.
This hamiltonian is rotationally invariant in spatial as well as in spin coordinates
since the spin-flip terms (i.e., with $\sigma'\neq \sigma$ )
arising from spin-orbit coupling are present in $H_{\text{int},dd}^{\text{HF1}}$.

Let us now comment on our choice for the basis set. Obviously, when all terms are present in
$H_{\text{int},dd}^{\text{HF1}}$, the results do not depend on the basis set. This may not be the case when
some matrix elements are omitted. For instance, in a common approximation, the terms involving
three and four different orbitals are neglected \cite{Anisimov93,Czyzyk94}. When using a basis set made of
spherical harmonics denoted by the value of the quantum number {\it m}, the three and four
orbital terms are a function of both B and C. Therefore if these terms are neglected without
changing the one and two different orbital matrix elements, the rotational invariance is destroyed
unless we set B=C=0, in which case the Coulomb type integrals $U_{mm'mm'}=A$ are completely
isotropic and the exchange integrals $U_{mm'm'm}$ vanish. This model is thus oversimplified
if we want to study spin and orbital magnetism. On the contrary when the basis set is built
from cubic harmonics, the three and four orbital matrix elements are a function of B only
\cite{Griffith61}. When these terms are neglected the rotational invariance is conserved
only if B is set equal to zero. Then in this model $U_{\lambda\mu\lambda\mu}=A+C$ and
$U_{\lambda\mu\mu\lambda}=C$ for any pair $\lambda, \mu$ of different {\it d} orbitals. This model
can describe correctly spin magnetism by an appropriate choice of the parameters but
cannot really account for orbital magnetism as shown in Ref.\onlinecite{Desjonqueres07}
on a simple model. This suggests to redefine the parameters determining the HF1 hamiltonian
as U, J, B with $U=(1/4)\sum_{\mu,\mu\ne\lambda} U_{\lambda \mu
\lambda \mu}$ and $J=(1/4)\sum_{\mu,\mu\ne\lambda} U_{\lambda \mu
\mu \lambda}$, the sums being independent of $\lambda$. These average values of Coulomb and
exchange interactions are given by $U=A-B+C$ and $J=5B/2+C$. Thus we expect that in the HF1
model the orbital magnetism, which is sensitive to the anisotropy of electronic interactions, is mainly
governed by B similarly to what is assumed in the orbital polarization ansatz. 

In the following we will consider also the HF2 model of Ref.\onlinecite{Desjonqueres07}
which is obtained from HF1 by setting B=0 and sometimes called the U,J model. Finally a
Stoner-like model called HF3 can be derived from HF1 under the following assumptions: i) the
net density matrix $\average{n_{i\lambda\mu}^{\sigma\sigma'}}$ is diagonal with elements equal to
the net occupation numbers $n_{i\lambda\sigma}$, ii) on each site {\it i} the exact $n_{i\lambda\sigma}$
are replaced by their average value $n_{i\sigma}=1/5 \sum_{\lambda}n_{i\lambda\sigma}$.
Then it is easily shown (see Appendix A) that the simplified hamiltonian can be written:

\begin{equation}
H_{\text{int},dd}^{\text{HF3}}=\sum_{i \lambda\sigma} (U_{\text{eff}} N_{i,d} -\sigma I_{dd}M_{i,d}/2 ) c_{i\lambda \sigma}^{\dag}  c_{i\lambda\sigma}.
\end{equation}

\noindent
where $U_{\text{eff}}=(9U-2J)/10$, $I_{dd}=(U+6J)/5$ is the $d$ Stoner parameter, $N_{i,d} (M_{i,d})$ are the net
{\it d} total population (spin momentum) at site {\it i} and $\sigma=+1(-1)$ for majority (minority) spin.
Note that the above conditions are
approximately obeyed for bulk transition metals but become questionable when the symmetry is 
lowered.

We must not forget that some electronic interactions are already included in $H_{TB}$ since
this hamiltonian has been parametrized by fitting LDA and GGA calculations. This is taken
into account following the treatment done in the "around mean field" LDA+U theory \cite{Czyzyk94}, i.e.,

\begin{equation}
\Delta H_{\text{int},dd}=H_{\text{int},dd}(n_{i\lambda\mu}^{\sigma\sigma'})-H_{\text{int},dd}({\bar
n}_i\delta_{\lambda\mu}\delta_{\sigma\sigma'})
\end{equation}

\noindent
with ${\bar n}_i=\sum_{\lambda\sigma}n_{i\lambda\sigma}/10$, whatever the model (HF1, HF2 or HF3).

Finally, the small exchange splittings of the {\it s} and {\it p} levels due to the spin 
polarization of {\it d} electrons is treated with a Stoner-like model and a Stoner parameter
$I_{sd}=I_{pd}=I_{dd}/10$, i.e.:

\begin{equation}
\Delta H_{\text{int},s(p)d}=-\sigma I_{s(p)d}M_{i,d}/2     
\end{equation}

\noindent
so that: 

\begin{equation}
\Delta H_{\text{int}}=\Delta H_{\text{int},sd}+\Delta H_{\text{int},pd}+\Delta H_{\text{int},dd}
\end{equation}

From the above discussion it is clear that HF2 differs from HF1 by
terms proportional to $B$, this is also true for HF3 as far as
this hamiltonian is justified. Eriksson {\sl et
al.}\cite{Eriksson90} have proposed to introduce an OPA term to
account for this difference. This term is written in meanfield $\Delta
E_{\text{OP}}=-\frac{1}{2}B \sum_i \average{\bm{L}_i}^2$ which
reduces to
 $-\frac{1}{2}B \sum_i \average{L_{iZ}}^2$ when the spin and orbital moment $\average{\bm{L}_i}$ are
collinear with the axis $Z$ which is verified along high symmetry directions (in the following
X, Y, Z will denote the framework in which Z is the magnetization direction).
 The corresponding hamiltonian is then:

\begin{equation}
H_{\text{OP}}=-B  \sum_{i \lambda \mu \sigma} \average{L_{iZ}} [L_{iZ}]_{\lambda \mu} c_{i \lambda \sigma}^{\dag} c_{i \mu \sigma}
\end{equation}

\noindent
where $[L_{iZ}]_{\lambda \mu}$ are the matrix elements of the
local orbital moment operator $L_{iZ}$. Note that in the basis of cubic harmonics
$[L_{iZ}]_{\lambda \mu}$ is not diagonal \cite{Autes06}. In the following we also compare
the results obtained with HF1 to those derived from HF2 or HF3 to which the OPA
term has been added.

As in our previous work \cite{Autes06} we use the HF3 model to determine the Stoner parameter $I_{dd}$ so
as to reproduce as closely as possible the variation of the bulk spin magnetic moment as a function
of the interatomic distance that can be obtained from a spin polarized DFT calculation. This
gives $I_{dd}=1.eV$. Finally from Fig.1 of the recent work by Solovyev \cite{Solovyev05} it
can be deduced that $U\simeq J$. As a consequence we have taken $U=J=5I_{dd}/7=0.71eV$ and similarly
to most previous works $B=0.14J$ \cite{Solovyev05}.

\section{Spin and orbital magnetism in bulk bcc iron}

Let us first consider the bulk bcc phase of Fe and compare the results obtained
with HF1 and HF3 using the TBGGA parameters. Indeed we do not expect strong
differences between HF2 and HF3 for the following reasons. First, in the absence
of the small perturbation due to spin-orbit coupling, the intra-atomic density 
matrix is diagonal for symmetry reasons in cubic crystals. Second, the populations
of the different orbitals with a given spin are rather similar in the bulk.

\begin{figure}[!fht]
\begin{center}
 \includegraphics*[scale=0.33,angle=0]{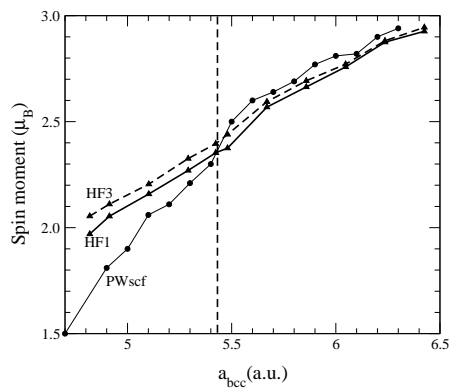}
\end{center}
\caption{Spin magnetic moment of bulk bcc iron as a function of the cubic lattice parameter from PWscf ab initio calculations,
HF1 and HF3 models.
The dashed straight line corresponds to the experimental equilibrium lattice parameter.}
\label{fig:Sz_bulk}
\end{figure}

\begin{figure}[!fht]
\begin{center}
 \includegraphics*[scale=0.33,angle=0]{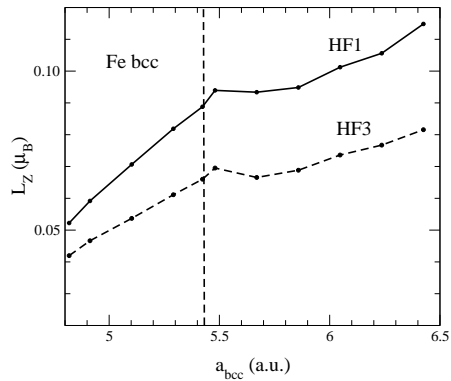}
\end{center}
\caption{Orbital magnetic moment of bulk bcc iron as a function of the cubic lattice parameter from the
HF1 and HF3 models. The dashed straight line corresponds to the experimental equilibrium lattice parameter.}
\label{fig:Lz_bulk}
\end{figure}

Calculations show that the equilibrium lattice parameter (5.35a.u.) is close to the experimental
value (5.43a.u.) and that the variation of the spin moment with the lattice parameter
is almost the same with the HF1 and HF3 models. For instance the spin moments are
2.34$\mu_B$ and 2.39$\mu_B$ with HF1 and HF3, respectively (see Fig.\ref{fig:Sz_bulk}) at
the experimental equilibrium lattice parameter.
On the opposite the orbital moment is significantly enhanced with the HF1 model
(see Fig.\ref{fig:Lz_bulk}) and in very good agreement with experiment 
(0.09$\mu_B$) \cite{Landolt86}.

\section{Spin and orbital magnetism at bcc iron Surfaces}

We have applied the HF1 and HF3 models to the study of the $(001)$ and
$(110)$ surfaces of bcc iron. The surfaces are modelled by slabs of 15 atomic layers which
is sufficient to avoid interactions between the two surfaces and to
recover the bulk  behaviour on the central layer. Interlayer relaxations are neglected in all
calculations and the interatomic distance is set equal to the experimental one 
($d_{\text{bulk}}^{\text{exp}}=4.69$a.u.). 
In such systems some atoms have a
reduced coordination compared to the bulk and therefore charge transfers are expected.
In order to avoid unphysically large charge transfers at the surface we have used
a penalty function (see Appendix B) which consists in adding to the energy functional a term  of the
type:

\begin{equation}
 E^{\text{pen}}_{\text{LCN}}=\lambda^{\text{pen}}_{\text{LCN}} \sum_i (\Delta q_i)^2 
 \end{equation}

\noindent
$\Delta q_i=q_i-q_0$ where $q_i$ is the Mulliken charge of site $i$,  $q_0$  the valence
charge and $\lambda^{\text{pen}}_{\text{LCN}}$ the penalization factor which must be taken large
enough to ensure local charge neutrality (in practice one takes $\lambda^{\text{pen}}_{\text{LCN}}=2.5$eV
which gives charge transfers below 0.1 electron per atom.).
The local charges $q_i$ are expressed as a function of the tight-binding expansion coefficients
$C_{i \lambda \sigma}^{n}$ of the eigenfunctions $n$ with respect to the atomic orbitals

\begin{equation}
 q_i=\text{Re}\Bigg( \sum_{\substack{\lambda j \mu  \sigma \\ n}} f_n C_{i \lambda \sigma}^{n \star} C_{j \mu \sigma}^{n} S_{ij}^{\lambda \mu} \Bigg) 
 \end{equation}

\noindent
where $f_n$ is the occupation factor and depends on the type of system. In periodic systems such as surfaces,
a broadening technique is used and $f_n=f(\varepsilon_n)$ is equal to the Fermi function at the 
energy level $\varepsilon_n$ for a given temperature.

 \begin{figure}[!fht]
\begin{center}
 \includegraphics*[scale=0.5,angle=0]{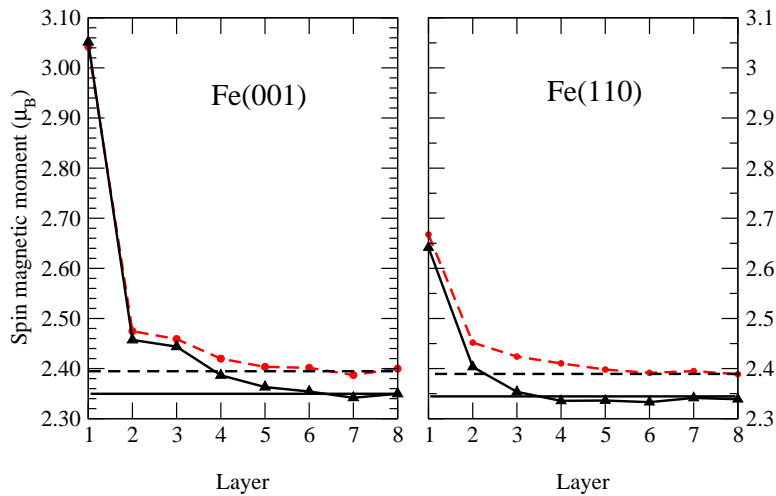}
\end{center}
\caption{Variation of the spin magnetic moment (per atom) on successive layers of $(001)$
and $(110)$ slabs (with 15 atomic layers) of bcc Fe obtained from HF1 (full lines) and HF3 (dashed lines) models. Layer
1 corresponds to the outermost layer and layer 8 to the central layer. The horizontal lines correspond to the bulk limit. }
\label{fig:Mz_surface_001_110}
\end{figure}

\begin{figure}[!fht]
\begin{center}
 \includegraphics*[scale=0.5,angle=0]{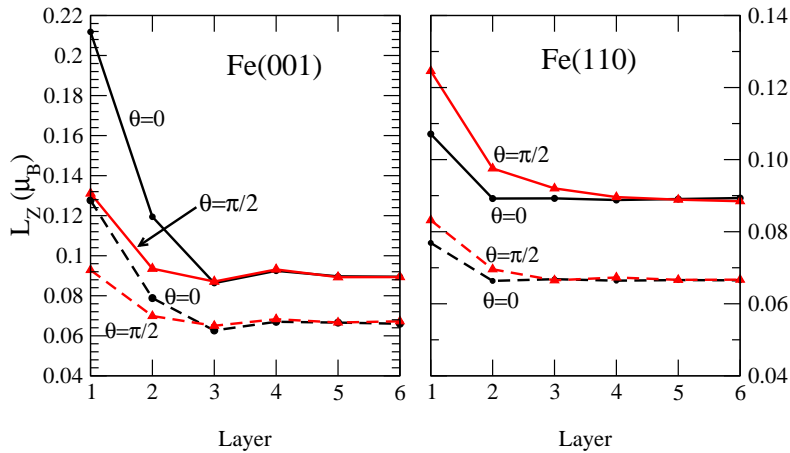}
\end{center}
\caption{Variation of the component of the local orbital magnetic moment on the
magnetization direction as a function of the atomic layer in the $(001)$
and $(110)$ slabs (15 layers) for a magnetization perpendicular ($\theta=0$)
and parallel ($\theta=\pi/2$) to the surface  from HF1 (full lines) and HF3 (dashed lines) models. 
For in-plane magnetization the dependence of the orbital moment on the orientation in the plane is negligible.}
\label{fig:Lz_surface_001_110}
\end{figure}

A standard iterative scheme is set up until input and output intra-atomic 
density matrix elements differ by less than $10^{-4}$ electron
per atom and the total energy of the slab does not change by more than $10^{-4}$eV.
We have calculated the spin and orbital moments decomposed on each atomic layer of the $(001)$ and
$(110)$ slabs. The results of our calculations are shown in Fig.\ref{fig:Mz_surface_001_110} and Fig.\ref{fig:Lz_surface_001_110}.
As expected the spin magnetic moment is enhanced in the vicinity of the surface,
this reinforcement being more pronounced for the $(001)$ for which the outermost layer is
almost saturated, than for the $(110)$ slab since the $(001)$ surface is more open than the $(110)$. 
The convergence to the bulk spin moment is also slightly faster for the $(110)$ slab than for
the $(001)$ slab. Finally, apart from a small shift of the bulk spin moment between HF1 (2.34$\mu_B$) 
and HF3 (2.39$\mu_B$) the two models  lead to very similar results. Let us also note that we have verified that the spin moment
is almost independent on the magnetization orientation.

At surfaces the orbital moment  is also strongly enhanced. However  in contrast
to the spin moment the type of model used is now crucial. Indeed the enhancement of the orbital magnetization
at the surface is much more pronounced with the HF1 model than with HF3. For magnetization perpendicular to
the surface one finds values of $\average{L_{iZ}}$  as large as 0.21$\mu_B$ on the outermost layer of 
the $(001)$ surface with the HF1 model, while it is 0.127$\mu_B$ with the HF3 model. The orbital
moment is therefore 2.40 larger at the surface than it is in the bulk (0.088$\mu_B$) with HF1 while it is slightly less than
twice larger with HF3. The orbital polarization effect is therefore amplified at the surface: 
a 63\% increase of the surface orbital moment  between HF3 and HF1 is obtained which must be compared to 33\% 
in the bulk (0.088$\mu_B$ with HF1 and 0.066$\mu_B$ with HF3). This general trend is
also observed on the $(110)$ surface though not so pronounced. 
Finally, contrary to the case of the spin moment the orbital moment $\average{L_{iZ}}$ depends
sensitively on the magnetization direction and moreover the two surfaces behave differently. It is found
that the orbital moment is noticeably larger for magnetization perpendicular to the surface
in the case of the $(100)$ slabs, while a slight increase of the orbital moment  is observed
for in-plane magnetization in the case of $(110)$ slabs. In that respect HF1 and HF3 models 
lead to very similar behaviors. Finally one should point out that for in-plane magnetization the
dependence of the orbital moment on the orientation in the plane is negligible.

\section{Magnetic properties of an iron monatomic wire}

In a preliminary work \cite{Desjonqueres07} we already investigated the orbital
contribution to the magnetic properties of the Fe monatomic wire and checked the
ability of the OPA to account for the existence of large orbital moments in 
one-dimensional structures by comparing to the results obtained from a full
Hartree-Fock decoupling of intra-atomic electronic interactions. To this aim we used a
simple tight-binding model in which only {\it d} electrons were taken into account.
However in this model only the self-consistent solution(s) of the hamiltonian
could be determined but not the total energy. Furthermore the number of
{\it d} electrons was fixed. On the contrary, the use of a {\it s, p, d} basis set
allows a charge redistribution between {\it sp} and {\it d} orbitals as
a function of the interatomic distance $d$ and the determination of the total
energy. In the following we present the results obtained for a monatomic Fe
wire with this realistic basis set using the TBLDA MP parameters for reasons
which have been discussed in Ref.\onlinecite{Autes06}. This will enable us
to confirm the qualitative trends obtained in the previous model and to
get quantitative results.

We have computed the total energy, the spin and orbital magnetic moments,
the band structure and the electronic transmission factor as a function of
the interatomic distance for the different models, i.e., HF1 and HF2 or HF3
without and with OPA. The results show that the equilibrium distance is
quite insensitive to the model chosen and is close to 4.05a.u. (2.14\AA)
with a slight dispersion smaller than 0.02a.u.. The spin magnetic moment
is quite independent on the direction of magnetization. This moment is also
almost unchanged when the OPA term is added to HF2 or HF3. Finally, while
the spin moment saturates between 3.7 and 3.8a.u. for HF1 and HF2, this
saturation occurs at a shorter distance (between 3.6 and 3.7a.u.) with HF3.
Note that these latter results are in much better agreement with ab initio
calculations (3.8a.u. \cite{Autes06}) than in the pure {\it d} band model.

\begin{figure}[!fht]
\begin{center}
 \includegraphics*[scale=0.5,angle=0]{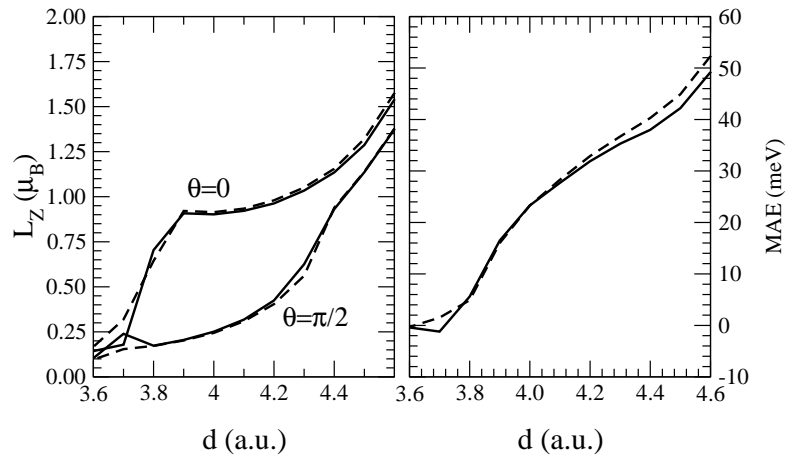}
\end{center}
\caption{Comparison between the results obtained with the HF2+OPA (full lines)
and HF3+OPA (dashed lines) hamiltonians for the monatomic Fe wire as a
function of interatomic distance: a) orbital moment for directions of magnetization
parallel ($\theta=0$)
and perpendicular ($\theta=\pi/2$) to the wire, b) the corresponding
magnetocrystalline anisotropy energy. }
\label{fig:HF2HF3}
\end{figure}

\begin{figure}[!fht]
\begin{center}
 \includegraphics*[scale=0.5,angle=0]{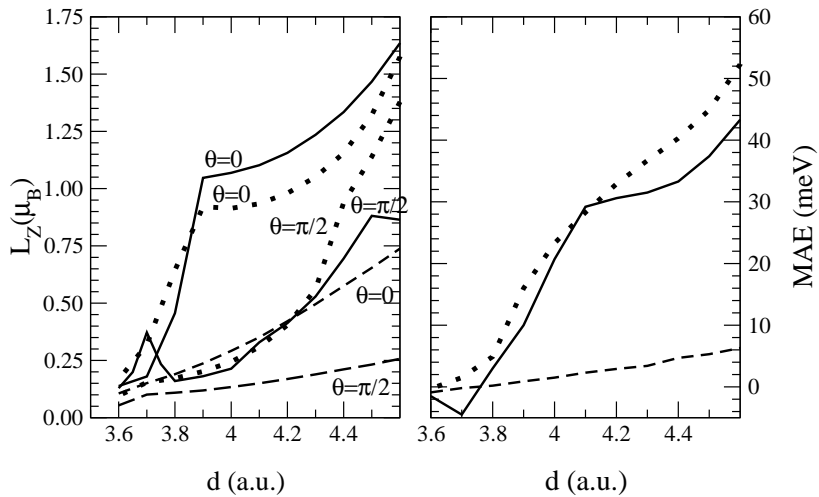}
\end{center}
\caption{ Same caption as Fig.\ref{fig:HF2HF3} but for the HF1 (full lines),
HF3 (dashed lines), HF3+OPA (dotted lines) hamiltonians. Note that the
curves referring to HF1 correspond to the most stable self-consistent 
solution for each value of $d$(see text).}
\label{fig:HF1HF3OPA}
\end{figure}

Let us now compare the orbital magnetic moments and MAE obtained with the various models. The models
HF2 and HF3 with (Fig.\ref{fig:HF2HF3}) or without OPA give very similar results save
for $3.6<d<3.8a.u.$, i.e., the domain of distances for which at least
one of these models leads to an unsaturated spin moment. As a consequence
we will now limit ourselves to the comparison between the HF1, HF3 and
HF3+OPA hamiltonians for magnetizations parallel ($\theta=0$) and
perpendicular ($\theta=\pi/2$) to the wire. A glance at Fig.\ref{fig:HF1HF3OPA} confirms,
as expected, that HF3 largely underestimates the orbital moment $\average{L_Z}$
for both magnetization directions, and the MAE. Adding the OPA to HF3
obviously increases these quantities, however, for the orbital moment
the difference with the result derived from HF1 does not keep the same sign
as a function of the interatomic distance and is rather large when the
spin is not saturated.

It is worthwhile to discuss the results given by HF1 in more details.
First at $\theta=0$ a sharp change of slope occurs for $\average{L_Z}$
at $d$=3.9a.u.. Let us recall \cite{Desjonqueres07} that at $\theta=0$
the wire eigenfunctions corresponding to the almost flat $\delta$ bands
located near the Fermi level are a linear combination of {\it d} spherical
harmonics (with {\it m}=2 or {\it m}=-2) centered at each site. When d increases
from 3.6a.u. the population of the highest energy band {\it m}=-2 decreases
and vanishes for $d\simeq 3.9a.u.$. This explains the change of
slope in the $\average{L_Z}$ curve. Note also that for $d>4.3a.u$, two
self-consistent solutions are found corresponding to an empty band
either with {\it m}=2 or {\it m}=-2 character and nearly opposite values of $\average{L_Z}$.
However the solution in which the {\it m}=-2 band is empty and $\average{L_Z}>0$
is always the most stable one. On the contrary at $\theta=\pi/2$ the
character of the eigenfunctions of the $\delta$ bands near the Fermi level
is $d_{xy}$ or $d_{x^2-y^2}$. Up to $d=4.a.u.$ these two bands are quasi
degenerate. Then, when $d\ge 4.1a.u.$ two self-consistent solutions are
found for which either $d_{xy}$ or $d_{x^2-y^2}$ are empty but, contrary 
to what occurs at $\theta=0$, they have almost the same total energy and
very similar values of $\average{L_Z}$. The change of slope at
$d=4.5a.u.$ corresponds to the switch of the ground state from one
case (empty $d_{x^2-y^2}$ band) to the other (empty $d_{xy}$ band).

It is interesting to compare our results with those of our previous works. In
Ref.\onlinecite{Autes06} we showed, using a Stoner like hamiltonian,
that the proportionality relation between the MAE and the anisotropy
of the orbital moment proposed by Bruno \cite{Bruno89}, by treating 
the spin-orbit coupling by means of perturbation theory, is almost
strictly verified when the interatomic distance in the wire is such 
that the spin moment is saturated. It is clear from Fig. \ref{fig:HF1HF3OPA} that this 
law is not satisfied with HF1 eventhough the MAE has the same sign as
$\average{L_Z(\theta=0)}-\average{L_Z(\theta=\pi/2)}$.

To illustrate the departure from the perturbation theory we have found useful to
study the variation of the MAE and orbital magnetization
with the angle between the monatomic wire and the magnetization axis. If one starts
a calculation from a non symmetric magnetic configuration (different from $\theta=0$ or
$\theta=\pi/2$) and let the self-consistent process iterate  until convergency it
should bring the system towards the easy axis. Therefore one has to find a way to follow
the evolution with respect to the angle $\theta$.
In order to constraint the angle $\theta$ between the spin moment and the $z$-axis we add
a penalty functional

\begin{equation}
E^{\text{pen}}_{\text{ang}}=\lambda^{\text{pen}}_{\text{ang}}(\cos \theta -\cos \theta_0)^2
\end{equation}

\noindent
to the total energy (see Appendix B). 

 \begin{figure}[!fht]
\begin{center}
\includegraphics*[scale=0.5,angle=0]{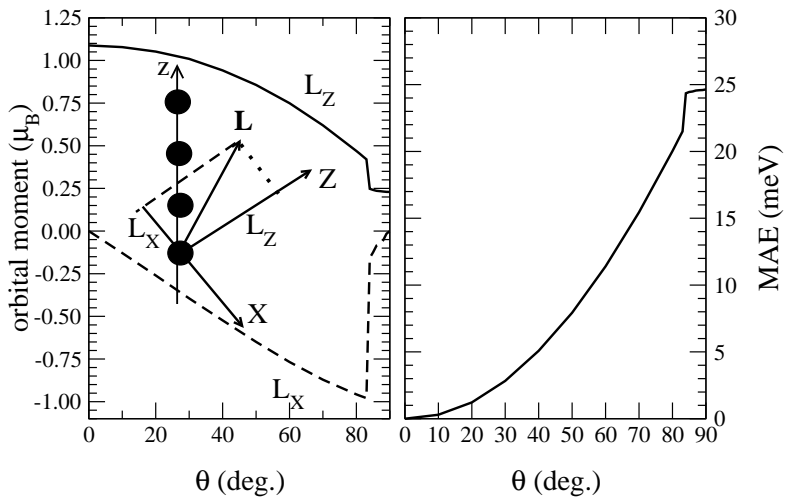}
\end{center}
\caption{a) Variation of the components of the orbital moment and of the magnetic anisotropy energy 
as a function of the direction of magnetization given by the angle $\theta$ for a monatomic wire
(interatomic distance $d=4.05$a.u.) obtained from HF1.}
\label{fig:HF1_theta}
\end{figure}

\noindent
In practice $\lambda^{\text{pen}}_{\text{ang}}=1$eV ensures that the angle $\theta$ does
not deviate by more than $0.1^{\circ}$ from $\theta_0$. Fig. \ref{fig:HF1_theta} shows the
results of our calculations with the HF1 model.
This clearly shows that the orbital moment and the MAE are strongly
modified when the anisotropy of the electronic interactions is taken
into account. The MAE and the orbital moment no longer follow
a simple $\sin^2\theta$  law and moreover present an abrupt variation
around $\theta=80^{\circ}$ which corresponds to an electronic transition. Note that
the position of this transition strongly depends on the interatomic distance. At larger
interatomic distances this transition occurs for smaller $\theta$ angles.

We must emphasize that the qualitative trends put forward
in Ref.\onlinecite{Desjonqueres07} are also obeyed when the basis
set is extended to include {\it s} and {\it p} orbitals, namely:
i) the orbital moment and MAE are largely underestimated with a
Stoner-like hamiltonian but they reach numerical values similar
to those observed experimentally in one dimensional systems when
HF1 is used, ii) adding the OPA to HF2 or HF3 leads to a fair
agreement with HF1 for both $\average{L_Z}$ and MAE when the spin
moment is saturated but this agreement deteriorates in relative
value for unsaturated spin polarization.

\begin{figure}[!fht]
\begin{center}
\includegraphics*[scale=0.5,angle=0]{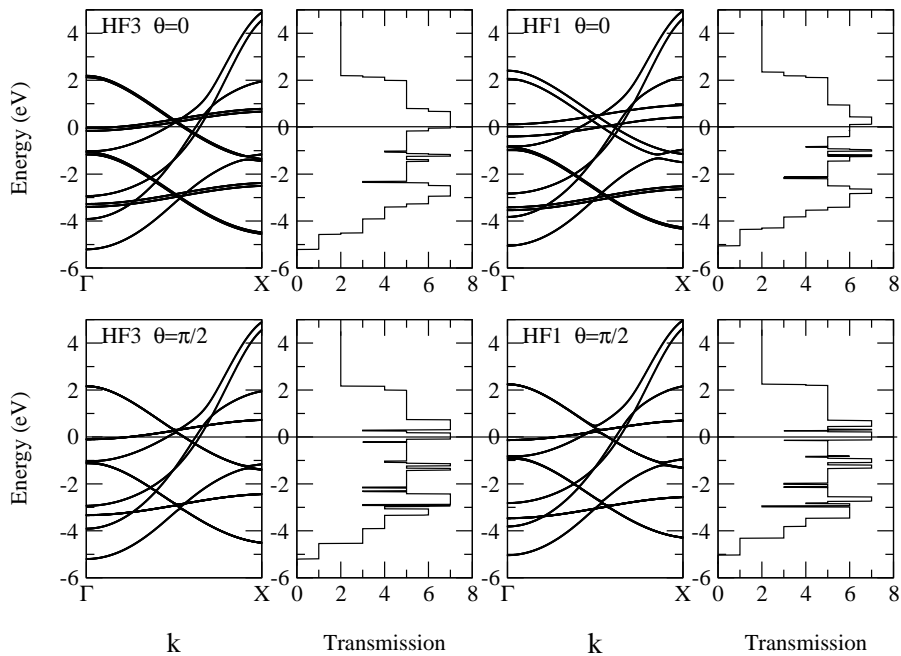}
\end{center}
\caption{Band structure and the corresponding electronic transmission factor of an iron monatomic
wire  for  magnetizations parallel ($\theta=0$)
and perpendicular ($\theta=\pi/2$) to the wire  from the HF1 and HF3 models. }
\label{fig:transmission_wire}
\end{figure}

Finally we show in Fig.\ref{fig:transmission_wire} the band structure obtained with HF1 and
HF3 for $\theta=0$ and $\theta=\pi/2$ at the equilibrium distance
from which the electronic transmission factor $T(E)$ is deduced by a
simple counting of the number of eigenfunctions at a given energy\cite{Databook}.
Some differences are observed, in particular at $\theta=0$: around
the Fermi level the energy domain inside which $T(E)=7$ is narrowed
when HF1 is used instead of HF3. Such differences should appear in more
complex geometries such as those obtained in the constriction region of
break junctions and could have important consequences on magnetoresistance
properties\cite{Viret06}.

\section{Magnetic properties of some iron clusters}

\begin{figure}[!fht]
\begin{center}
 \includegraphics*[scale=0.33,angle=0]{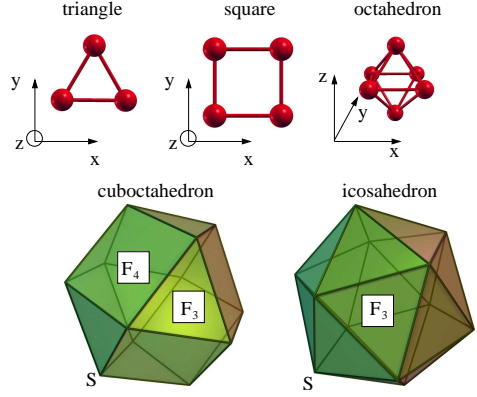}
\end{center}
\caption{Geometry of the iron clusters. In the case of cuboctahedral and icosahedral
geometries each apex $S$ and the centre of the cluster are occupied by an iron atom.
$F_4$ and $F_3$ denote the center of a square and triangular facet, respectively. }
\label{fig:clusters}
\end{figure}

In the previous sections we have studied 3- (bulk), 2- (surface), and 1- (wire) dimensional periodic systems.
To end this work we have investigated zero-dimensional structures, i.e., unsupported clusters.
We have considered  five clusters with various geometries presented in Fig. \ref{fig:clusters}.

Three of these clusters (triangle, square and regular octahedron)
are made of geometrically equivalent atoms,  while the two others (cuboctahedron and
icosahedron) are built from a central atom surrounded by twelve  atoms forming
the outer shell. Since it was shown in the previous section that HF2 and HF3 basically give
the same results for saturated systems we have restricted our study to the three models: HF1, HF3 and HF3+OPA.
In a first step we have minimized the total energy with respect to the nearest neighbour distance.
We only consider an homogenous contraction of the cluster and ignore Jahn Teller distorsions.
Note that the discrete levels of the clusters have been filled with electrons up to the
highest occupied level (HOMO), i.e., the occupation factor $f_n=1$ or $0$ except when
the HOMO level is degenerate and not completely filled. In the latter case the different wave functions
corresponding to the HOMO level have been equally populated in order to preserve the
cluster symmetry. Similarly to the case of surfaces a penalization function has been 
used to avoid large charge transfers (see Appendix B).

\begin{figure}[!fht]
\begin{center}
\includegraphics*[scale=0.5,angle=0]{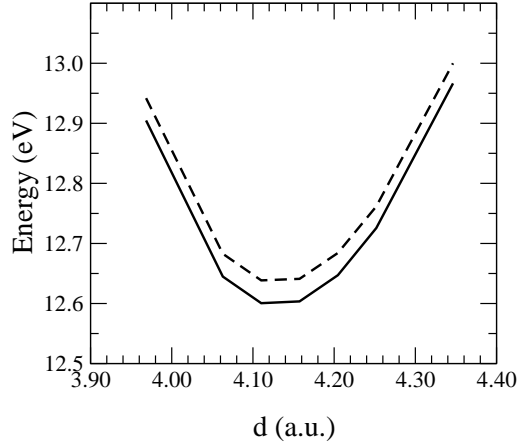}
\end{center}
\caption{Total energy as a function of the interatomic distance of a triangular iron cluster
from HF1 (full line) and HF3 (dashed line) models. Note that the HF3+OPA curve would be undistinguishable from the HF3 one.}
\label{fig:energy_triangle_d}
\end{figure}

As illustrated in Fig. \ref{fig:energy_triangle_d} the equilibrium interatomic distance is almost insensitive to the type of 
hamiltonian. We also checked that the magnetization orientation does not influence the
equilibrium distance either. As a consequence the relaxed structures have been determined from HF3 calculations and
a magnetization along a high symmetry direction (see Fig. \ref{fig:clusters}).
The results are summarized in Table \ref{tab:eq_cluster}. For all clusters
a contraction of the equilibrium bond-length with respect to the calculated bulk one ($d_{\text{bulk}}=4.63$a.u.) is
obtained and the general trend stating that the contraction 
decreases with the average coordination is well obeyed.
The interatomic distance of the triangle and square is contracted by 11\%,
the octahedron by 7\% and the cuboctahedron by 3\% with respect to the bulk. In the
case of the icosahedron one should distinguish between the radial $d_r$ and intrashell $d_t$
nearest neighbour distance, the latter being  about 5\% larger than the former
($d_t=1.051d_r$). At equilibrium one finds that the average value $(d_t+d_r)/2=4.53$a.u.
is very close to the interatomic distance of the cuboctahedron.

\begin{table}[!fht]
\begin{tabular}{|c|c|c|c|c|c|c|}
\hline
            &   triangle &  square & octahedron & cuboctahedron & icosahedron \\
	    \hline
$d_{eq}$(a.u.) &  4.14      &   4.16  &  4.31      &  4.50         &  4.42    \\
\hline          
\end{tabular}
\caption{Equilibrium first nearest neighbour distance of various iron clusters from
HF3 calculations. Note that these values are almost identical to those
obtained with HF1 calculations, and are almost independent on the magnetization
axis. In the case of the icosahedron $d_{eq}$ is the radial first nearest neighbour distance $d_r$. }
\label{tab:eq_cluster}
\end{table}

We have then carried out a detailed study of the MAE and
of spin and orbital moments for these equilibrium structures with the three
hamiltonians. The spin moment is almost insensitive to the hamiltonian
used. We also found that the spin moment is carried by $d$ electrons
and is almost saturated. Consequently the $d$ spin moment $M_{i,d}$ on a given site $i$ can approximately be
obtained from the number of $d$ electron $N_{i,d}$ by the relation  $M_{i,d}=10-N_{i,d}$. 
This simple rule is rather well obeyed 
for all clusters (see Table \ref{tab:spin_moment_cluster}) save for the cuboctahedron and 
icosahedron in which the central atom is depleted in $d$ electrons and is not saturated. 
Since it is well known from theoretical works that fcc iron has a tendency to form
complicated magnetic structures we have tried  to start the self-consistent scheme 
for the cuboctahedron and icosahedron from an antiferromagnetic magnetic configuration.
We were not able to find any antiferromagnetic solution except when allowing
charge transfers by setting $\lambda^{\text{pen}}_{\text{LCN}}$ to zero. In that case a rather large charge
transfer is obtained (especially in the case of the icosahedron)
from the inner atom to the outershell and one can find a self-consistent solution
where the central atom has a spin magnetic moment opposite to that of outershell atoms.
The ferromagnetic solution however remains the most stable solution.

\begin{table}[!fht]
\begin{tabular}{|c|c|c|c|c|c|c|}
\hline
            &   triangle &  square & octahedron & cuboctahedron & icosahedron \\
	    \hline
$N_{i,d}$             &   7.21     &   7.24  &   7.31      &  6.80/7.14         &   6.90/7.12    \\
$M_{i,d}$             &   2.68     &   2.57  &   2.44      &  2.42/2.63         &   2.28/2.63    \\
$10-N_{i,d}$          &   2.79     &   2.76  &   2.69      &  3.20/2.86         &   3.10/2.88    \\   
$M_i$  &   2.66     &   2.50  &   2.33      &  2.47/2.62         &   2.33/2.63    \\
\hline          
\end{tabular}
\caption{Number of $d$ electrons $N_{i,d}$, $d$ and total spin moment $M_{i,d}$ and $M_i$ (in $\mu_B$)on each site of the five
clusters. Note that these quantities  are almost independent on the model, for instance, the spin moments 
do not differ by more than 0.01 $\mu_B$. The two values for the cuboctahedral and icosahedral clusters
refer to the central and peripheral atoms, respectively.}
\label{tab:spin_moment_cluster}
\end{table}

\begin{figure}[!fht]
\begin{center}
\includegraphics*[scale=0.5,angle=0]{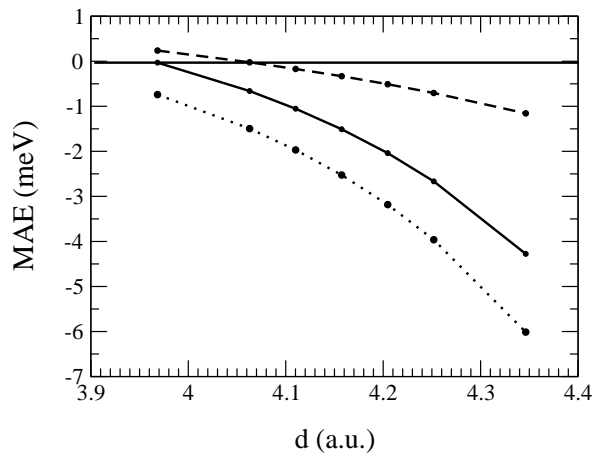}
\end{center}
\caption{Magnetic anisotropy energy (MAE) $E_{100}-E_{001}$ as a function of the interatomic distance 
of a triangular iron cluster from HF1 (full line), HF3 (dashed line) and HF3+OPA (dotted line) models. 
Note that the MAE curve $E_{010}-E_{001}$ would be
undistinguishable from $E_{100}-E_{001}$ since there is almost no in-plane anisotropy.}
\label{fig:MAE_triangle_d}
\end{figure}

Let us now discuss the MAE.  First, it is worth noting that the MAE
is quite sensitive to the interatomic distance. For illustration
we have calculated the MAE of a triangular iron cluster 
(see Fig. \ref{fig:MAE_triangle_d}) as a function of the interatomic distance.
As expected the HF3 model systematically leads to smaller MAE than the HF1 or  HF3+OPA model.
 Moreover the MAE may change sign when the interatomic distance increases as seen in Fig. \ref{fig:MAE_triangle_d}
where the easy axis switches from out of plane (positive MAE) to in plane (negative MAE) at 
$d=4.06$a.u. with the HF3 model. It is also found that the in-plane anisotropy is extremely small
and of the order of some hundredth of meV. The results are given in Table \ref{tab:MAE}
for the five clusters at equilibrium. The behavior of 
the square cluster is very different from the triangular one since
the three models agree to find an out-of-plane easy axis and a strong in-plane anisotropy
(the diagonal axis being strongly unfavorable) but the numerical values are very dependent on the
model (see Table \ref{tab:MAE}). 
The octahedron is the only cluster for which the three models agree to predict an absence
of anisotropy (smaller than $2.10^{-5}$eV).
Finally the comparison of the cuboctahedron and icosahedron is instructive since 
these two structures only differ by small atomic displacements \cite{Barreteau00b}. 
The cuboctahedron shows a rather large anisotropy 
in favor of the direction denoted as $F_3$, while the icosahedron shows a very 
weak anisotropy due to its more spherical shape.
For both structures, the addition of the OPA term to HF3 does not lead to a significant
improvement compared to HF1.

Since non-collinear magnetism is always a possible issue in clusters, we have
checked for the five clusters that the most stable spin configuration is the ferromagnetic collinear one.
To this aim we have initialized our calculation by  arbitrary non-collinear magnetic configurations
and iterated until convergency. In most cases a ferromagnetic collinear configuration is obtained. However,
depending on the initial conditions, several collinear and non-collinear antiferromagnetic-like
 configurations, ({\sl i.e.,} with zero total magnetic moment) are 
sometimes found, but their energy is always above the ferromagnetic one.
Interestingly when the calculation converges towards the ferromagnetic configuration 
the self-consistent scheme proceeds as follows: in the first iterations the magnetic
moments tend to align along a given direction (depending on the initial condition), then
once the moments are almost aligned it takes a long time for the system to converge towards
the easy axis. For all clusters we were able to find the easy axis by this procedure except for the
cuboctahedron. In fact depending on the initial conditions the calculation sometimes converges towards
the $S$ (apex) configuration, sometimes towards the $F_3$ (center of triangular facet), meaning that
the total energy possesses several local minima.

 \begin{table}[!fht]
\begin{tabular}{|c|c|c|}
\hline
                     \multicolumn{3}{|c|}{triangle}  \\
\hline		     
                &    \multicolumn{2}{|c|}{MAE=$E_{\bm u}-E_{001}$ (meV)}  \\
       \hline
$\bm u \rightarrow$       &      $(100)$  &   $(010)$  \\
\hline
HF1             &    -1.264      &   -1.270         \\
HF3             &    -0.246      &   -0.246          \\
HF3+OPA         &    -2.225      &   -2.253         \\
\hline
                     \multicolumn{3}{|c|}{square}  \\
\hline		     
                &    \multicolumn{2}{|c|}{MAE=$E_{\bm u}-E_{001}$}  \\
       \hline
$\bm u \rightarrow$        &  (100)          &  $(110)$             \\
\hline
HF1             &  8.911	 &  15.245	   \\
HF3             &  5.378	 &  6.3190   \\
HF3+OPA         &  20.560        &  26.346   \\
\hline
                     \multicolumn{3}{|c|}{octahedron}  \\
\hline		     
                &    \multicolumn{2}{|c|}{MAE=$E_{\bm u}-E_{001}$ (meV)}  \\
       \hline
$\bm u \rightarrow$         &  $(100)$       &    \\
\hline
HF1             &  -0.004        &		  \\
HF3             &  -0.001	 &		   \\
HF3+OPA         &   0.021	 &		  \\
\hline
                     \multicolumn{3}{|c|}{cuboctahedron}  \\
\hline		     
                &    \multicolumn{2}{|c|}{MAE=$E_{\bm u}-E_{F_4}$ (meV)}  \\
       \hline
$\bm u \rightarrow$         &  $S$               &  $F_3$  \\
\hline
HF1             &  -1.676	     &  -2.208 	\\
HF3             &  -0.331	     &  -0.441	\\
HF3+OPA         &  -0.358            &  -0.359 \\
\hline
                     \multicolumn{3}{|c|}{icosahedron}  \\
\hline		     
                &    \multicolumn{2}{|c|}{MAE=$E_{\bm u}-E_S$ (meV)}  \\
       \hline
$\bm u \rightarrow$         &    $F_3$             &    \\
\hline
HF1             &  0.114        &	      \\
HF3             &  0.020        &	       \\
HF3+OPA         & -0.020        &	       \\
\hline
\end{tabular}
\caption{ Magnetic anisotropy energy in meV for various iron clusters from HF1, HF3 and HF3+OPA
models calculated at the equilibrium interatomic distance. For the cuboctahedron and the icosahedron
the direction $\bm{u}$ is determined by the vector joining the central atom to an apex (S) and to the centers of
a triangular facet ($F_3$) or a square facet ($F_4$)} 
\label{tab:MAE}
\end{table}

Let us now consider the orbital moment. The results of our
calculations are presented in Tables \ref{tab:Lz1} and \ref{tab:Lz2}.
It is seen that the orbital moment is very
sensitive to the magnetization orientation as at surfaces and in the monatomic wire. 
Furthermore two structurally equivalent atoms
now become "magnetically" inequivalent and can bear
orbital moments that differ by a factor of more than two. For instance in
a cuboctahedron with a magnetization pointing towards the center of the square
facet $F_4$ the 8 atoms forming the upper and lower square facet (with respect to
the magnetization axis) have an orbital moment about twice smaller than that of the
four other outershell atoms whatever the model. 
The orbital moment of the central atom is close to the bulk bcc value.
In contrast the icosahedron shows a more equally distributed orbital moment and the
orbital moment on the central atom is significantly larger than in the bulk. Moreover the difference 
between the values obtained from  the various hamiltonians is  much less pronounced.

Finally, as expected and contrary to the spin moment, the type of hamiltonian
has most often a strong influence on the numerical value of the orbital moment.
We find that HF3 systematically predicts the smallest ones. However, contrary to
the case of MAE, there is now
a fairly good agreement between HF1 and HF3+OPA.
It should be pointed out that when two magnetic orientations  
have almost the same energy the total orbital moment in these two directions are
almost identical. This can be verified from Tables \ref{tab:MAE} and \ref{tab:Lz1} 
in the case of the $(100)$ and $(010)$
directions of the triangle. One should  however not conclude that in the HF1 model
the MAE is simply related
to the orbital moment. Indeed, with this model, the easy axis of the triangle corresponds
to the highest orbital moment but this is not true for the square.

 \begin{table}[!fht]
\begin{tabular}{|c|c|c|c|}
\hline
                     \multicolumn{4}{|c|}{triangle}  \\
\hline		     
                &    \multicolumn{3}{|c|}{$\average{L_Z}$ ($\mu_B$)}  \\
       \hline
$\bm Z \rightarrow$         &    $(001)$      &   $(100)$      &   $(010)$           \\
\hline
HF1             &   0.133 (3)     &   0.177 (2)    &  0.216 (2)       \\
                &                 &   0.236 (1)    &  0.159 (1)       \\
HF3             &   0.090 (3)     &   0.112 (2)    &  0.129 (2)        \\
                &                 &   0.138 (1)    &  0.104 (1)        \\
HF3+OPA         &   0.130 (3)     &   0.189 (2)    &  0.229 (2)        \\
                &                 &   0.238 (1)    &  0.159 (1)        \\
\hline
                     \multicolumn{4}{|c|}{square}  \\
\hline		     
                &    \multicolumn{3}{|c|}{$\average{L_Z}$ ($\mu_B$)}  \\
       \hline
$\bm Z \rightarrow$         &    $(001)$      &   $(100)$         &   $(110)$           \\
\hline
HF1             &  0.325 (4)       &  0.337 (4)       &  0.234 (2)          \\
                &                  &                  &  0.259 (2)          \\
HF3             &  0.255 (4)       &  0.164 (4)       &  0.133 (2)            \\
                &                  &                  &  0.142 (2)            \\
HF3+OPA         &  0.487 (4)       &  0.362 (4)       &  0.249 (2)            \\    
                &                  &                  &  0.270 (2)            \\
\hline
                     \multicolumn{4}{|c|}{octahedron}  \\
\hline		     
                &    \multicolumn{3}{|c|}{$\average{L_Z}$ ($\mu_B$)}  \\
       \hline
$\bm Z \rightarrow$         &    $(001)$      &   $(100)$       &            \\
\hline
HF1             &    0.124 (4) &  0.128 (4)   &              \\
                &    0.133 (2) &  0.124 (2)   &              \\
HF3             &    0.087 (4) &  0.094 (4)   &              \\
                &    0.101 (2) &  0.087 (2)   &              \\
HF3+OPA         &    0.128 (4) &  0.140 (4)   &              \\
                &    0.153 (2) &  0.128 (2)   &              \\		
\hline
\end{tabular}
\caption{Local orbital moments of iron triangular, square and octahedral clusters for various
magnetization orientations $Z$(see Fig. \ref{fig:clusters}), from HF1, HF3 and HF3+OPA
hamiltonians at the equilibrium distance. The number of sites having the same orbital
moment for symmetry reasons is given in parenthesis. In the case of the $(110)$ direction of the square
the largest value corresponds to the diagonal of the square perpendicular to the magnetization.} \label{tab:Lz1}
\end{table}

 \begin{table}[!fht]
\begin{tabular}{|c|c|c|c|}
\hline
                     \multicolumn{4}{|c|}{cuboctahedron}  \\
\hline		     
                &    \multicolumn{3}{|c|}{$\average{L_Z}$ ($\mu_B$)}  \\
       \hline
$\bm Z \rightarrow$         &    $F_4$      &   $S$         &   $F_3$           \\
\hline
HF1             &  0.098 (1)  &  0.101 (1)      &   0.102       (1)      \\
                &  0.156 (8)  &  0.246 (8)      &   $\approx$0.23 (12)  \\
                &  0.304 (4)  &  0.153 (2)      &              \\
	        &             &  0.182 (2)      &              \\	

HF3             &  0.063 (1)  &  0.063 (1)      &   0.064        (1)       \\
                &  0.098 (8)  &  0.143 (8)      &   $\approx$0.13  (12)          \\
                &  0.182 (4)  &  0.095 (2)      &              \\
	        &             &  0.104 (2)      &              \\

HF3+OPA        &  0.103 (1)   &  0.105 (1)      &   0.105        (1)     \\
               &  0.158 (8)   &  0.255 (8)      &   $\approx$0.23  (12)    \\
               &  0.321 (4)   &  0.162 (2)      &                \\
               &              &  0.179 (2)      &                \\	       

\hline
                     \multicolumn{4}{|c|}{icosahedron}  \\
\hline		     
                &    \multicolumn{3}{|c|}{$\average{L_Z}$ ($\mu_B$)}  \\
       \hline
$\bm Z \rightarrow$         &      $S$            &      $F_3$           &              \\
\hline
HF1             &  0.202   (1)     &   0.202  (1)    &              \\
                &  0.317   (10)    &  $\approx$0.32(12) &              \\
                &  0.304   (2)     &                    &              \\		
HF3             &  0.207   (1)     &   0.207  (1)    &              \\
                &  0.222   (10)    &  $\approx$0.22(12) &              \\
                &  0.210   (2)     &                    &              \\		
HF3+OPA         &  0.271   (1)     &  0.271  (1)     &              \\
                &  0.312   (10)    &  $\approx$0.28 (6) &              \\
                &  0.268   (2)     &  $\approx$0.32 (6) &              \\		
\hline
\end{tabular}
\caption{Local orbital moments of iron for cuboctahedral and icosahedral clusters for various
magnetization orientations $Z$ determined by the vector joining the central atom to an apex (S) and 
the centers of a triangular facet ($F_3$) or of a square facet ($F_4$), 
from HF1, HF3 and HF3+OPA hamiltonians at the equilibrium inter-atomic distance. 
The number of sites having the same orbital
moment for symmetry reasons is given in parenthesis, except for the direction denoted as $F_3$
for which the orbital moments are very similar on all the external atoms.
In the case of the $S$ direction of the cuboctahedron the smallest orbital moment in the outershell
corresponds to the two atoms with a binding direction parallel to the magnetization.} \label{tab:Lz2}
\end{table}

\section{Conclusion}

In conclusion, this work presents an extended study of orbital polarization effects
on the magnetic properties of iron systems with various dimensionalities using
a $s, p, d,$ tight-binding hamiltonian with spin-orbit coupling and $dd$ electronic
interactions treated in the Hartree-Fock scheme at different levels of approximations.
The results obtained from the full Hartree-Fock interaction hamiltonian HF1, i.e.,
involving all matrix elements of the Coulomb interactions with their exact orbital
dependence as a function of the three Racah parameters, have been systematically
compared with those of simplified hamiltonians including or not the orbital
polarization ansatz (OPA). As expected it is found that the one-parameter Stoner-like 
hamiltonian HF3 which, similarly to the L(S)DA approach, neglects the orbital dependence
of Coulomb interactions, leads to reasonable values of the spin moment but largely
underestimates the orbital moment and the magnetocrystalline anisotropy energy (MAE).
The two-parameter (U, J) HF2 model, in which three and four
orbital matrix elements are neglected and those with two different orbitals are
averaged in the cubic harmonics basis, leads practically to the same results as
the Stoner-like hamiltonian, at least when the spin moment is saturated. In both
these simplified hamiltonians the addition of the OPA, which is not really justified
on theoretical grounds, gives largely improved values of the orbital moment but
is much less reliable for the MAE, at least in clusters or when the spin moment
is not saturated. With the full interaction hamiltonian HF1, the orbital
moment and the MAE attain numerical values of the same order of magnitude as those
measured experimentally although larger, which is not surprising since the experiments
have been performed on supported clusters or chains.

Furthermore HF1 is fully rotationally invariant as well in spatial and in spin
coordinates. The influence of spin-flip terms is rather weak in the case of iron
but is expected to increase drastically with the spin-orbit coupling parameter. This
could be important for platinum which has been shown to be magnetic in monatomic
wires \cite{Delin04}. Finally it should be emphasized that it is not
much more computer demanding to deal with the HF1 model. It is thus of prime importance
to work with this model for the study of orbital magnetism and MAE in low dimensional systems either in a realistic
 $s, p, d,$ tight-binding basis set or to implement it in ab-initio codes
of the LDA+U type.   

\appendix

\section{The relation between HF1 and HF3 interaction hamiltonians}

In this appendix, we show, using the real $d$ orbital basis set, that the
Stoner-like hamiltonian HF3 can be obtained from the full Hartree-Fock interaction hamiltonian
$H_{\text{int},dd}^{\text{HF1}}$ with the following approximations: 

i) The intra-atomic matrix density is assumed to be diagonal with respect to
both orbital and spin indices.

ii) For each spin and at each site the exact population of the $\nu\sigma$ spin-orbital 
is replaced by its average value, i.e., $n_{i\sigma}=1/5 \sum_{\nu}n_{i\nu\sigma}$
then the matrix elements of the approximate electronic interaction hamiltonian become:

\begin{equation}
H_{i\lambda\sigma,i\mu\sigma}=n_{i\sigma}\sum_{\nu}(U_{\nu\lambda\nu\mu}-U_{\nu\lambda\mu\nu})
+n_{i-\sigma}\sum_{\nu}U_{\nu\lambda\nu\mu}
\end{equation}

It can be shown \cite{Griffith61} that the matrix elements $U_{\lambda\mu\nu\eta}$ vanish
when three indices are equal and that the elements involving three different orbitals
obey the relations $\sum_{\nu}U_{\nu\lambda\nu\mu}=0$ and $\sum_{\nu}U_{\nu\lambda\mu\nu}=0$.
Consequently the approximate hamiltonian is diagonal and: 

\begin{equation}
H_{i\lambda\sigma,i\lambda\sigma}=n_{i\sigma}\sum_{\nu}(U_{\nu\lambda\nu\lambda}-U_{\nu\lambda\lambda\nu})
+n_{i-\sigma}\sum_{\nu}U_{\nu\lambda\nu\lambda}
\end{equation}

\noindent
By replacing $n_{i\sigma}$ by $(N_{i,d}+\sigma M_{i,d})/10$ we find:

\begin{equation}
H_{i\lambda\sigma,i\lambda\sigma}=\sum_{\nu}(U_{\nu\lambda\nu\lambda}-\frac{U_{\nu\lambda\lambda\nu}}{2})
\frac{N_{i,d}}{5}-\sigma\sum_{\nu}U_{\nu\lambda\lambda\nu}\frac{M_{i,d}}{10}
\end{equation}

\noindent Carrying out the summations over $\nu$ leads to:

\begin{equation}
H_{i\lambda\sigma,i\lambda\sigma}=\frac{9U-2J}{10}N_{i,d}-\frac{\sigma}{2}(\frac{U+6J}{5})M_{i,d}
\end{equation}

\noindent
which are the matrix elements of the electronic interaction hamiltonian in the HF3 model.

\section{The penalty functions}

The usual way to implement constraints in a variational problem is to 
use a Lagrange-multiplier formalism. For example the normalization constraint of the wave 
functions leads to the eigenvalue
problem, the Lagrange-multiplier being the eigenvalues.
However this approach is not always very well suited to more general 
constraints. A very useful
approach is to add a supplementary term to the total energy function\cite{GebauerPhD,Cohen04}. 
This term called
penalty function is equal to zero when the constraint is fulfilled and 
large and positive
when it is not. In the present work we have used 
a local charge neutrality (LCN)
penalty $E^{\text{pen}}_{\text{LCN}}$ and a $\theta$ angle penalty 
$E^{\text{pen}}_{\text{ang}}$.
We have taken the following forms for these penalty functions:
 
\begin{equation}
E^{\text{pen}}_{\text{LCN}}=\lambda^{\text{pen}}_{\text{LCN}} \sum_i(\Delta q_i)^2
\end{equation}

\noindent
and:

\begin{equation}
E^{\text{pen}}_{\text{ang}}=\lambda^{\text{pen}}_{\text{ang}} \sum_i(\cos \theta_i -\cos \theta_{i 0})^2
\end{equation}

\noindent
$\Delta q_i=q_i-q_0$ where $q_i$ is the Mulliken charge of site $i$,  
$q_0$  the valence
charge and $\lambda^{\text{pen}}_{\text{LCN}}$ the penalization factor 
which must be taken large
enough to ensure local charge neutrality; 
$\theta_i$ is the local magnetization angle on atom $i$ and 
$\theta_{i 0}$ the angle that
one wants to impose. Using the expression of $q_i$ and $\cos 
\theta_i=M_{i z}/\sqrt{M_{i x}^2+ M_{i y}^2+M_{i z}^2}$  in terms
of the tight-binding expansion coefficients of the wave functions, 
the calculation of the derivative of the 
penalized total energy 
with respect to these expansion coefficients leads to an eigenvalue 
problem where the hamiltonian
is now modified by an additional "renormalizing" term which matrix 
elements reads:

\[  \lambda^{\text{pen}}_{\text{LCN}} (\Delta q_i + \Delta 
q_j)\sigma_{0}  S_{ij}^{\lambda \mu}  \]

\noindent
and

\[ -\lambda^{\text{pen}}_{\text{ang}}  
(\bm{B}^{\text{pen}}_i+\bm{B}^{\text{pen}}_j).\bm{\sigma}  
S_{ij}^{\lambda \mu} \]

\noindent
in the case of LCN and angle penalization, respectively, where 
$\sigma_{0}$ is the identity matrix and
$\bm{\sigma}$ the Pauli matrix vectors. $\bm{B}^{\text{pen}}_i$ is an 
effective constraining magnetic field
that exerts a torque to bring the magnetization along the $\theta_{i 0}$ 
axis which expression is:

\begin{equation}
\bm{B}^{\text{pen}}_i= \lambda^{\text{pen}}_{\text{ang}}
\frac{(\cos \theta_i -\cos \theta_{i 0})}{(M_{i x}^2+ M_{i y}^2+M_{i 
z}^2)^{3/2}} \left (
\begin{array}{c}
M_{i x} M_{i z} \\
M_{i y} M_{i z} \\
-M_{i x}^2-M_{i y}^2
\end{array}\right)
\end{equation}

\noindent
In practice $\lambda^{\text{pen}}_{\text{LCN}}=2.5$eV and 
$\lambda^{\text{pen}}_{\text{ang}}=1$eV
ensure charge transfers below 0.1 electron per atom and deviations from 
the angle $\theta_{i0}$ smaller than $0.1^{\circ}$.

The modified eigenvalue problem has to be solved iteratively until 
convergence of the Mulliken
charges (and total energy). Finally it should be noted that if one 
writes the total energy as a sum
of the occupied eigenvalues, one should not forget to subtract an additional 
double counting term arising from the penalization function.

\end{document}